\documentclass[journal=apchd5,manuscript=letter, layout=traditional]{achemso}
\setkeys{acs}{articletitle=true}
\setkeys{acs}{keywords = true}
\usepackage{graphicx}
\usepackage[version=3]{mhchem}
\usepackage{graphicx,siunitx,microtype}
\author{Luis A. Guerra Hern\'andez}
\affiliation{Centro At\'omico Bariloche and Instituto Balseiro, Comisi\'on Nacional de Energ\'ia At\'omica (CNEA) - Universidad Nacional de Cuyo (UNCUYO), 8400 Bariloche, Argentina.}
\affiliation{Instituto de Nanociencia y Nanotecnolog\'ia (INN-Bariloche), Consejo Nacional de Investigaciones Cient\'ificas y T\'ecnicas (CONICET), Argentina.}
\author{Andr\'es A. Reynoso}
\affiliation{Centro At\'omico Bariloche and Instituto Balseiro, Comisi\'on Nacional de Energ\'ia At\'omica (CNEA) - Universidad Nacional de Cuyo (UNCUYO), 8400 Bariloche, Argentina.}
\affiliation{Instituto de Nanociencia y Nanotecnolog\'ia (INN-Bariloche), Consejo Nacional de Investigaciones Cient\'ificas y T\'ecnicas (CONICET), Argentina.}
\affiliation{Departamento de F\'isica Aplicada II, Universidad de Sevilla, E-41012 Sevilla, Spain}
\author{Alejandro Fainstein}
\email{afains@cab.cnea.gov.ar}
\affiliation{Centro At\'omico Bariloche and Instituto Balseiro, Comisi\'on Nacional de Energ\'ia At\'omica (CNEA) - Universidad Nacional de Cuyo (UNCUYO), 8400 Bariloche, Argentina.}
\affiliation{Instituto de Nanociencia y Nanotecnolog\'ia (INN-Bariloche), Consejo Nacional de Investigaciones Cient\'ificas y T\'ecnicas (CONICET), Argentina.}
\DeclareGraphicsExtensions{{.pdf},{.png}}

\title{Has the chemical contribution a secondary role in SERS?}
\begin{document}
\begin{abstract}
It is an established understanding that the electromagnetic contribution (the plasmon-mediated enhancement of the laser and scattered local electromagnetic fields) is the main actor in Surface Enhanced Raman Scattering (SERS), with the so-called chemical (molecule-related) contribution assuming only, if any, a supporting role. The conclusion of our comprehensive resonant study of a broad range of nanosphere lithography based metallic substrates, with covalently attached 4-mercaptobenzoic acid monolayers used as probe (standard molecules which are non-resonant in solution), is that this accepted understanding needs to be revised. We present a detailed resonant SERS study of Metal-film over nanosphere (MFON) substrates which is done {\em both} by scanning the laser wavelength, and by tuning the plasmon response through the nanosphere diameter which is varied from 500 to 900 nm. Far and local field properties are characterized through measures of optical reflectivity and SERS efficiency, respectively, and are supported by numerical simulations. We demonstrate that the SERS efficiency depends indeed on the electromagnetic mechanism, determined by the plasmonic response of the system, but we observe that it is also strongly defined by a chemical resonant contribution related to a metal-to-ligand electronic transition of the covalently bound probe molecule. Optimum amplification occurs when the plasmon modes intersect with the ligand-to-metal chemical resonance, contributing synergically both mechanisms together. Quite notably, however, the largest SERS signal is tuned with the metal-to-ligand transition, and typically does not follow the wavelength dependence of the plasmon modes when varying the nanosphere size. The same general trend is observed for other nanosphere lithography based substrates, including sphere-segment void cavities and hexagonally ordered triangular nanoparticles, using both Ag or Au as the plasmonic metal, and also with a commercial substrate (Klarite).
Interestingly, this extensive comparative investigation shows in addition that metal-film-over-nanosphere substrates are significantly better than the rest in terms of Raman efficiency and homogeneity. We conclude that a deep understanding of both the electromagnetic and chemical mechanisms is necessary to fully exploit these substrates for analytical applications.
\end{abstract}
%=================================
%=================================

\section{Motivation}
%~\\[1em] % keywords too close to text

The plasmonic properties of metal nanostructures are of great interest since they exhibit localized surface plasmons resonances (LSPR), with electromagnetic fields located on the nanostructured surface, and with resonance energy depending on the material, size and shape of the nanostructure. The interaction between photons impinging from the far field and the LSPR leads to a confinement of the electromagnetic field, enhancing its magnitude and opening the door to plasmon enhanced optical spectroscopies, including surface-enhanced Raman spectroscopy (SERS)~\cite{Moskovits1985,Schatz2006,LeRu2008,Novotny2011}.The enhancement of either or both the laser and Raman scattered fields by the plasmon resonances is identified as the electromagnetic contribution to SERS~\cite{Moskovits1985,Schatz2006,LeRu2008,Hao2004,Xu2003,Moskovits2005}. Raman scattering can also be amplified through electronic resonances either intrinsic to the probed molecule (if electronic resonant transitions of the molecule are available at the selected laser excitation), or related to the specific chemical interaction between the molecule and the supporting substrate (a situation that can play a role particularly for molecules that are not intrinsically resonant at the selected laser energy). A variety of different mechanisms have been envisaged involving this latter type of electronic resonances, mostly involving metal-to-ligand (ML) transitions of the bound molecule, and are generally grouped in what is called the chemical contribution to SERS.~\cite{Persson1981,Adrian1982,Lombardi1986,Campion1998,Otto2002,Otto2005,Persson2006,Lombardi2012,Londero2013,Lombardi2017} Maybe in part because of this complexity, and specificity for different systems, but also because it is assumed to be much weaker and of lesser relevance than the electromagnetic plasmonic enhancement (although some reports position it in the $10^5-10^7$ enhancement range)~\cite{Zhao2006,Fromm2006}, the so-called chemical contribution has remained in the backstage, at best assuming a secondary supporting role in SERS. An enormous collection of experimental and theoretical work seems to support this view, which has lead through the design of proper plasmonic substrates to great progress, with successful applications that range from ultrasensitive analytical methods to single-molecule spectroscopies~\cite{Kneipp1997,LeRu2006}, including the optical monitoring of single-molecule single-electron transport~\cite{Cortes2010}.

To be fair, however, very few of the reported experimental investigations treat in depth the issue of the resonant enhancement in SERS. Reportedly, less than one percent of the published works address this problem.~\cite{Haynes2003}  The reasons are simple to understand. To do such a resonant investigation either the excitation wavelength needs to be tuned with enough flexibility through the relevant wavelength range,~\cite{Kurouski2017,Zuloaga2011,McFarland2005,Tognalli2012} or the nanostructures must allow for a reproducible tuning of the plasmonic resonances across the available laser source wavelength.~\cite{Haynes2003,Jensen1999,Hulteen1999,Weimer2001,Jackson2004,Benz2016} Laser wavelength scans are rarely performed because they require the availability of a large set of lasers or widely tunable sources, and access to a triple spectrometer for efficient stray light rejection. Plasmon scans, in turn, require a flexible and controlled tunability of the selected technology, and typically imply an important load of fabrication, structural and optical characterization, and experimental work. Different groups have in any case pursued either one of these strategies, and this has been done mostly interpreting the results in the framework of the electromagnetic enhancement mechanism, broadly confirming its relevance in the observed amplification. Notwithstanding this general agreement, it is also interesting to note that the emerging phenomenology is not universal, nor simple. Painstaking experiments based on an extensive number of nanosphere lithography nanoparticle substrates of varying resonant energy, indicate that a correlation between the plasmon resonance and the fixed laser wavelength indeed exists, although with a seemingly noisy correlation.~\cite{Haynes2003} Some other experiments that report on precisely the same kind of substrates, but scanning the laser wavelength, evidence what seems to be a clear {\it blue shift} of the Raman resonance respect to the plasmon extinction maximum.~\cite{McFarland2005} This was interpreted as the convoluted resonant effect of the incoming laser and the (red-shifted) out-going Stokes Raman scattered fields. Other reports observe the reverse. That is a {\it red shift} of the Raman resonance respect to the plasmon extinction maximum.~\cite{Kurouski2017} This contrasting result has been interpreted as a supposedly universal phenomena related to the dissipation-induced red-shift of the local field modeled as due to the metal charges oscillating as a forced oscillator.~\cite{Zuloaga2011} In one case were the SERS resonant enhancement was studied with great detail for individual nanoparticles on a mirror, the overall intensity was shown to increase with the particle size and not when the plasmon resonance matched the excitation laser.~\cite{Benz2016} Notwithstanding this rather complex landscape with contrasting evidence, when looked in detail, it is clear that the general agreement is that the problem is understood in its essence, with the electromagnetic enhancement being the main character of the plot~\cite{Moskovits2005}, and only some voices raised arguing that such explanation is not complete.~\cite{Lombardi2012}

To illuminate this already extensively studied subject which, in our view, has nevertheless remained rather obscure, we present what is in our understanding the first comprehensive resonant SERS study in which experiments are carried out simultaneously at a wide variety of excitation wavelengths and also tuning the surface plasmon resonance by controlling the SERS substrate structure. This is done on substrates fabricated with nanosphere lithography~\cite{Hulteen1995,Haynes2001} which are extensively studied and well known for their reproducibility and homogeneity. We describe first the resonant wavelength plasmonic properties of metal-film over nanosphere (MFON) substrates\cite{Dick2002,Baia2006} with M=Au and Ag, which are studied through optical reflectivity, SERS measurements, and numerical simulations. The resonant wavelength of the localized plasmons in these MFON substrates can be adjusted by changing the diameter of the nanosphere, with wavelengths varying in the NIR-VIS spectral range ($\sim$400-1100~nm). For the SERS studies, a self-assembled monolayer of 4-mercaptobenzoic acid (4-MBA) was used as probably the most extensively used Raman probe that forms an ordered and densely packed monolayer on the surface of the metal.~\cite{Michota2002,Cortes2009}  This Raman molecule in solution displays electron transitions in the UV. That is, it is non-resonant in the spectral range where the plasmons reside. Previous investigations have shown, however, that this and similar thiol-bound molecules develop a resonant electronic ligand-to-metal transition (L-M $\sim$700~nm) when covalently attached to either Au or Ag.~\cite{Demuth1981,Tognalli2011,Morton2009,Valley2013} Notably, we find that the maximum SERS efficiency does not follow the plasmon dispersion but, on the contrary, is observed when a plasmon mode becomes resonant with the ligand-to-metal electronic resonance. We extend this investigation to other ordered plasmonic substrates, namely Au segment sphere void cavities~\cite{Tognalli2012,Kelf2006,Mahajan2007,Mahajan2009} and triangular nanoparticles~\cite{Jensen1999} both fabricated by nanosphere lithography, and a commercial Au substrate composed of square pyramidal pits (Klarite)~\cite{Perney2007}, with coincident results. Our conclusion is that SERS is a single effect drawing on plasmon and metal-to-ligand resonances which are intimately tied to each other and cannot straightforwardly be considered separately. While this does not contradict the established conviction on the relevance of the electromagnetic amplification, since determining the true SERS enhancement factor is critical for analytical determinations,~\cite{LeRu2007} it becomes clear that the more complex unified perspective will need to be considered for real world applications.~\cite{Lombardi2012,Lombardi2017,Johansson2005,Xu2004}

%%%=================================
%%%=================================
\section{Samples and experimental set-up}
%=================================
%=================================

\subsection{Fabrication of periodic hexagonal arrays}
Polystyrene spheres dispersed in a 1$\%$ water solution (Duke Scientific) are introduced between a glass substrate covered with a 100~nm Au-film (\textit{Platypus}) and a clean glass slide, with a separation of about 300~$\mu$m. The Au-film is immersed in a 1~mM cysteamine ethanolic solution overnight to enhance the polystyrene spheres adsorption. During the drying process in an incubation chamber, a sweeping meniscus forms along the substrate, pulling the spheres towards the substrate into a close-packed hexagonal monolayer. To fabricate the MFON substrates this self-assembly of the template is followed by a physical vapor deposition of Au or Ag (Vega and Camji S.A.I.C). The Au and Ag-films were vapor-deposited over 500-900~nm polystyrene spheres, in a homemade evaporator, which achieves a residual pressure of $\sim$1$\times$10$^{-7}$~torr. The control of the material fusion temperature is done manually, with a current of 10-15~A. The triangular lattice of ordered nanoparticles is obtained starting from 50~nm thick Au-FON substrates, and after removal of the latex spheres by sonication in acetone, isopropanol and Milli-Q water. Segment void sphere cavity substrates are fabricated following the same procedure of deposition of the polystyrene nanospheres into a close-packed hexagonal monolayer, followed by electrochemical deposition of Au from a Gold salt solution (TG-25 RTU, Technic Inc.), with a deposition rate of 2.5mC/min, followed by removal of the plystyrene spheres by sonication in a sequence of solvents.~\cite{Tognalli2011,GuerraHernandez2015} The studied square pyramidal pit substrates were acquired from the company Klarite.

%=================================
%=================================
\subsection{Substrate characterization and resonant SERS experiments}
Reflectivity measurements were used to identify the plasmon modes of all the different studied substrate arrays, using a fully automated Wollam WVASE32 variable angle spectroscopic ellipsometer with focusing probes, presenting a 100~$\mu$m circular spot on the sample with a numerical aperture of $\sim$0.02. SEM images were recorded with a FEI field-emission gun, Nova NANO-SEM 230, operating at 10~kV and with a tilt angle of 45$^\circ$. Surface roughness was characterized also in all the substrates through atomic force microscopy (AFM) with an AFM Veeco Dimension 3100, with MESP tip. Signal homogeneity was monitored with a LabRam HR Evolution Raman microscope using the He-Ne 633~nm laser line (close to the M-L resonance at $\sim 675$~nm) and taking 400 spectra in a $5 \times 5 \mu$m$^2$ square area with a x100 microscope objective of NA=0.9. SERS measurements were performed using a triple-stage Raman spectrometer (Horiba Jobin-Yvon T64000) operating in subtractive mode, and equipped with a liquid nitrogen-cooled charge-coupled device (CCD). The excitation was performed using the 514, 568, 647 and 676~nm lines of an Ar-Kr laser and a continuously tunable Ti-Sapph laser between 680 and 780~nm. The position on the sample was manually controlled, and the Raman signals were collected in a backscattering configuration with a collection lens of focal length +10~cm. The entrance slit of the spectrometer was kept at 200~$\mu$m, as the Raman peaks for the metal-adsorbed molecule are relatively broad and do not require high spectral resolution. Typical acquisition times were from 5 to 10~s, depending on the wavelength and the sample. Typically 10 spectra were acquired for each wavelength and substrate at different positions, with the average providing the SERS intensity and the statistical dispersion the shown error bars. 

%%%=================================
%%%=================================
\section{Results and Discussion}
%=================================
%=================================
\subsection{Plasmonics modes and SERS in Au-film over nanosphere (AuFON) substrates}

\begin{figure}[hbt!]
\centering
\includegraphics[width=0.8\textwidth]{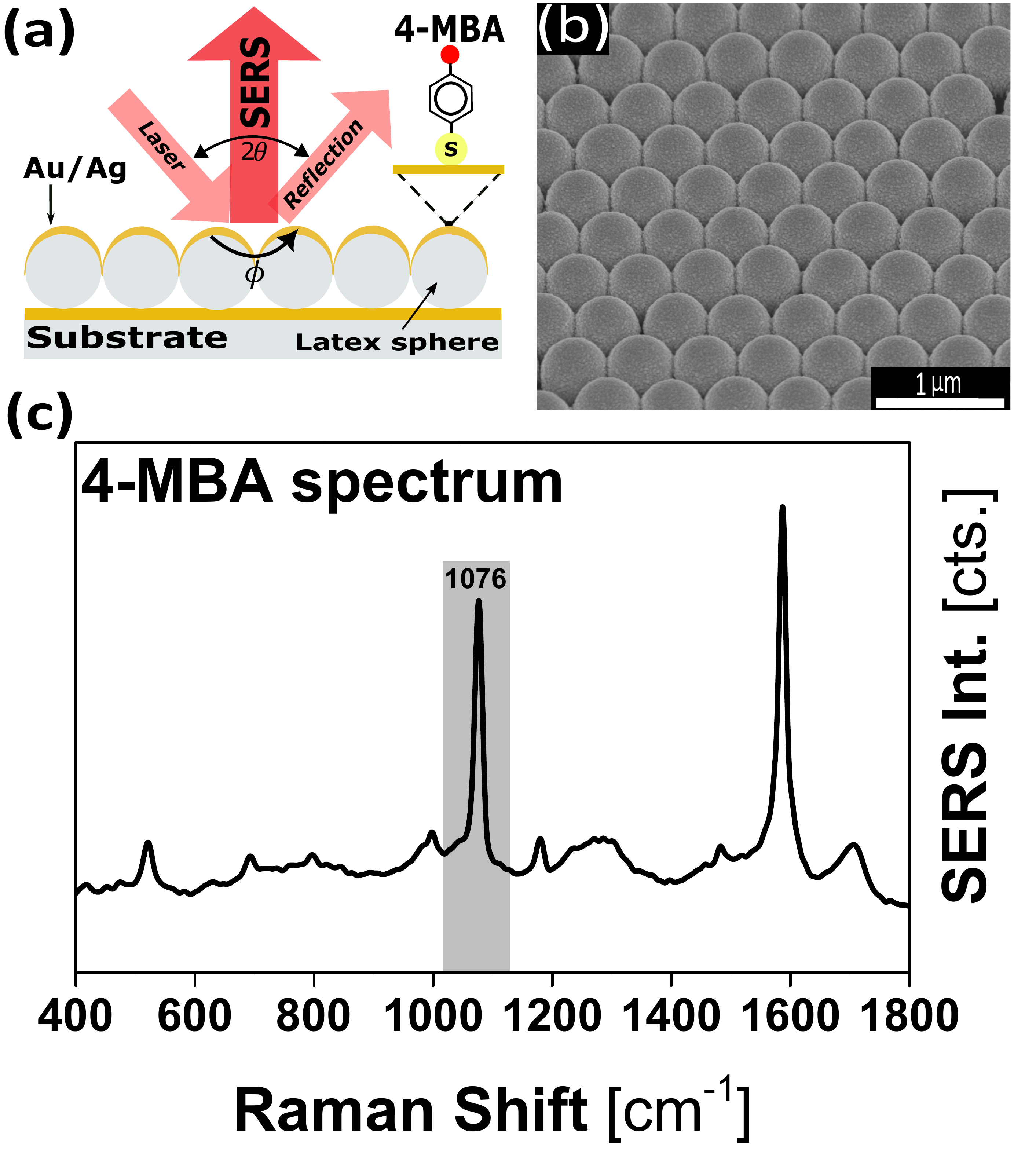}
\caption{(a) Schematic of the 4-MBA SERS experiment on metal-film on nanosphere (MFON) substrates. (b) High-resolution SEM of a typical nanostructure AuFON of 500~nm in diameter of polystyrene sphere, tilt=45$^\circ$. (c) Typical spectrum of 4-MBA absorbed in AuFON. The Raman mode at 1076~cm$^{-1}$ used to monitor the SERS efficiency is highlighted.
}
\label{Fig01}
\end{figure}

Figure~\ref{Fig01}(a) presents a scheme of the AuFON substrate and the experimental configuration with light incident at a small angle, and scattered light collected along the substrate normal. Panel (b) in this same figure shows a high-resolution SEM image of a typical AuFON fabricated with 500~nm diameter polystyrene spheres, taken with a tilt angle of 45$^\circ$. The close-packed and highly ordered formation of the support nanosphere mask can be clearly identified. A typical  spectrum of 4-MBA immobilized on AuFON is displayed in Fig.~\ref{Fig01}(c), with the aromatic-ring breathing vibration used to monitor the SERS intensity in the following sections highlighted with grey background at 1076~cm$^{-1}$. 

\begin{figure}[hbt!]
\centering
\includegraphics[width=0.8\textwidth]{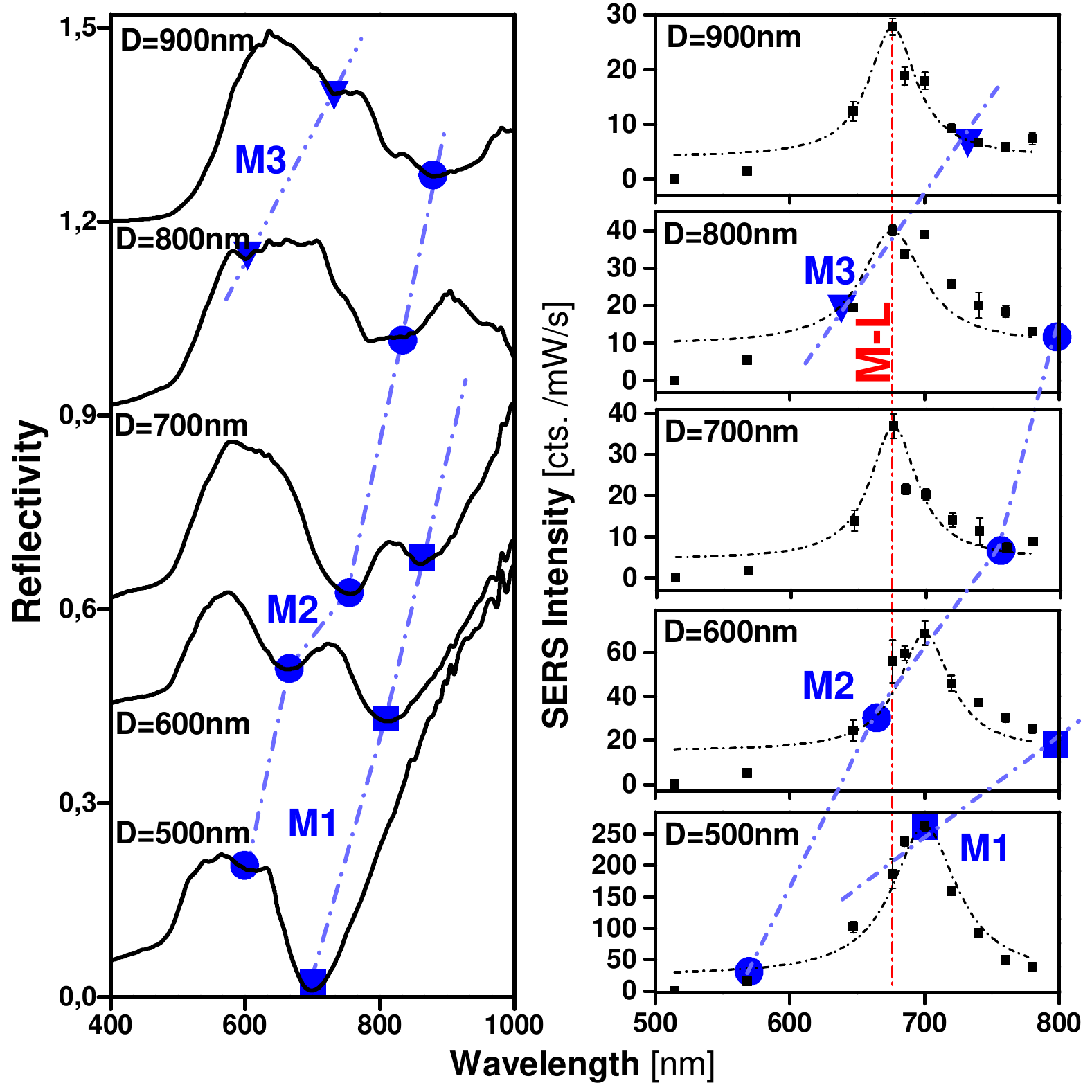}
\caption{Left: Reflectivity of AuFON substrates with varying diameter of the polystyrene spheres. Curves are vertically offset by steps of 0.3 for clarity. The dispersion of the plasmon modes (M1, M2 and M2) is indicated with blue symbols and guides to the eye. The experiments were done with TE polarization, and incident angle of 25$^\circ$.
Right: Raman intensity as a function of laser wavelength for the same substrates presented in the left panel. The spectra were also acquired with parallel TE polarizations. The shown Gaussian curves are guides to the eye. The symbols and connecting lines identify the plasmon resonances determined by the reflectivity measurements shown in the left panel. The vertical line signals the metal-to-ligand transition (M-L).
}\label{Fig02}
\end{figure}

The far-field response of the plasmon resonances in the studied AuFON substrates is shown in Fig.~\ref{Fig02} (left panel), for Au polystyrene nanosphere diameters in the range 500-900~nm. The reflectivity experiments were performed with TE polarization and light incident at an angle of 25$^\circ$, with wavelengths between 400-1000~nm. Each curve is vertically offset by steps of 0.3 for clarity. We note that for $\lambda<$550~nm the reflectivity decreases due to the interband transitions characteristic of Au. Three resonant absorptions can be identified in the wavelength range of 600-900~nm, indicated by blue triangles, circles and squares, respectively. These absorptions are associated with LSPR plasmons of the AuFON structures, and which we label as M1, M2 and M3 for increasing energy (decreasing wavelength). We have observed that  these modes do not vary significantly with the polar ($\theta$) and azimuthal ($\phi$) angles. Importantly, as expected the resonance wavelength of these modes strongly blue-shifts when reducing the diameter of the polystyrene spheres.

%=================================
%=================================
The resonant SERS study for the different AuFON substrates is shown in Fig.~\ref{Fig02}(right). The SERS intensity was monitored for the different samples through the amplitude of the 1076~cm$^{-1}$ Raman peak. Each point is the average of ten measurements at different spots, with the error bars indicating their dispersion. Raman intensities are given in counts per mW.s, with all spectra acquired exactly under the same experimental conditions. In Fig.\ref{Fig02}(right) the vertical red line indicates the wavelength of the ligand-to-metal transition (L-M), while the blue symbols are guides to the eye identifying the plasmon resonances obtained from the reflectivity curves in Fig.\ref{Fig02}(left). Two aspects can be highlighted from these results. First, irrespective of the AuFON period, the most intense Raman signals are detected in all studied cases very close to the M-L transition, with only a small red-shift of the resonance maxima when the M1 plasmon mode approaches this resonance. Second, the maximum detected signal augments when a plasmon mode is close to the M-L transition, and this is particularly noteworthy for the M1 mode which leads to the strongest SERS (bottom spectrum in Fig.\ref{Fig02}(right)). 

\subsection{Plasmonics modes and SERS in Ag-film over nanosphere (AgFON) substrates}

\begin{figure}[hbt!]
\centering
\includegraphics[width=0.8\textwidth]{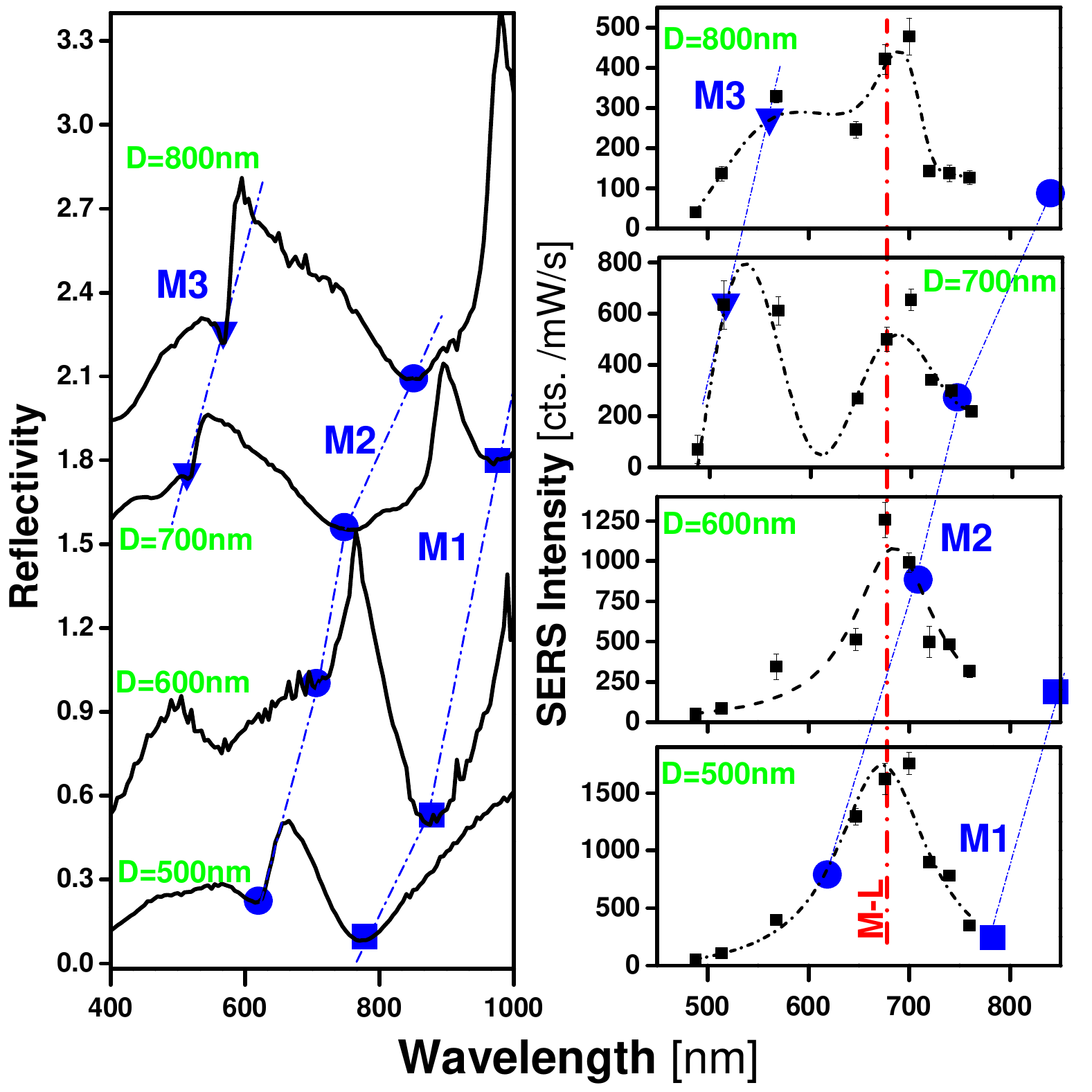}
\caption{Left: Reflectivity of AgFON substrates with varying diameter of the polystyrene spheres. Curves are vertically offset by steps of 0.3 for clarity. The dispersion of the plasmon modes (M1, M2 and M2) is indicated with blue symbols and guides to the eye. The experiments were done with TE polarization, and incident angle of 25$^\circ$.
Right: Raman intensity as a function of laser wavelength for the same substrates presented in the left panel. The spectra were also acquired with parallel TE polarizations. The shown black dash-dotted curves are guides to the eye. The symbols and connecting lines identify the plasmon resonances determined by the reflectivity measurements shown in the left panel. The vertical line signals the metal-to-ligand transition (M-L).
}\label{Fig03}
\end{figure}

The results presented in Fig.\ref{Fig02} indicate that plasmons (that is, the electromagnetic enhancement) play an important role on the SERS efficiency, but also suggest that this might not be separable from the metal-to-ligand resonant contribution (that is, the so-called chemical enhancement). To provide additional data on this phenomena we present in Fig.\ref{Fig03} a similar investigation but now performed on Ag instead of Au MFON substrates. Again the left panel in Fig.\ref{Fig03} presents the reflectivity measurements, while the right panel shows the corresponding SERS intensity curves for AgFON substrates made with polystyrene NPs of sizes ranging from 500 to 800nm. Similarly to the Au case, three plasmon resonances can be identified that blue-shift with decreasing NP size (labeled as M1, M2 and M3 in the figure). Several aspects of the SERS resonant curves can be mentioned. First, again Raman resonant maxima are observed for all substrates at the same wavelength coincident with the M-L transition, irrespective of the specific pattern of plasmon modes of the substrate. The absolute value of the scattering efficiency increases when a plasmon mode is close to the M-L transition, but the resonant peak does not follow the plasmon dispersion: it is fixed at the spectral position of the M-L transition. Again as for the Au case the proximity of the M1 plasmon mode seems to be specially relevant to provide the largest scattering efficiencies (bottom spectrum in Fig.\ref{Fig03}(right). Second, for the Ag film on nanosphere substrates a second maximum can be identified, in this case following the M3 plasmon mode (top two spectra in Fig.\ref{Fig03}(right)). This difference  when compared with the AuFON substrates can be traced to the inter-band transitions existent in Au and absent in Ag,~\cite{Johnson1972} which are expected to quench the Raman resonances due to absorption at wavelengths below $\sim 600$nm. Note that part of the intensity of the M3 resonance when compared to that due to the M1 mode can be ascribed to the $\omega^4$ enhancement which between 700 and 550~nm varies by a factor of 2.5.~\cite{LeRu2006CPL} Third, as is standard in SERS Ag provides significant larger enhancement than Au when structurally similar substrates are compared (at least a factor of 5 comparing Figs.~\ref{Fig02} and \ref{Fig03}).

Overall then, the results for AgFON substrates in Fig.\ref{Fig03} confirm the main conclusions obtained from those made from Au in Fig.\ref{Fig02}, namely: the electromagnetic plasmon-mediated enhancement is clearly part of the menu, but the chemical contribution has no secondary role in this phenomena, at least for the studied MFON substrates and for a quite universally used covalently attached probe molecule. As we will see next, these conclusions are rather general and valid for a much larger set of plasmonic substrates. 

\subsection{Comparative study between different ordered plasmonic substrates}

\begin{figure}[hbt!]
\centering
\includegraphics[width=0.8\textwidth]{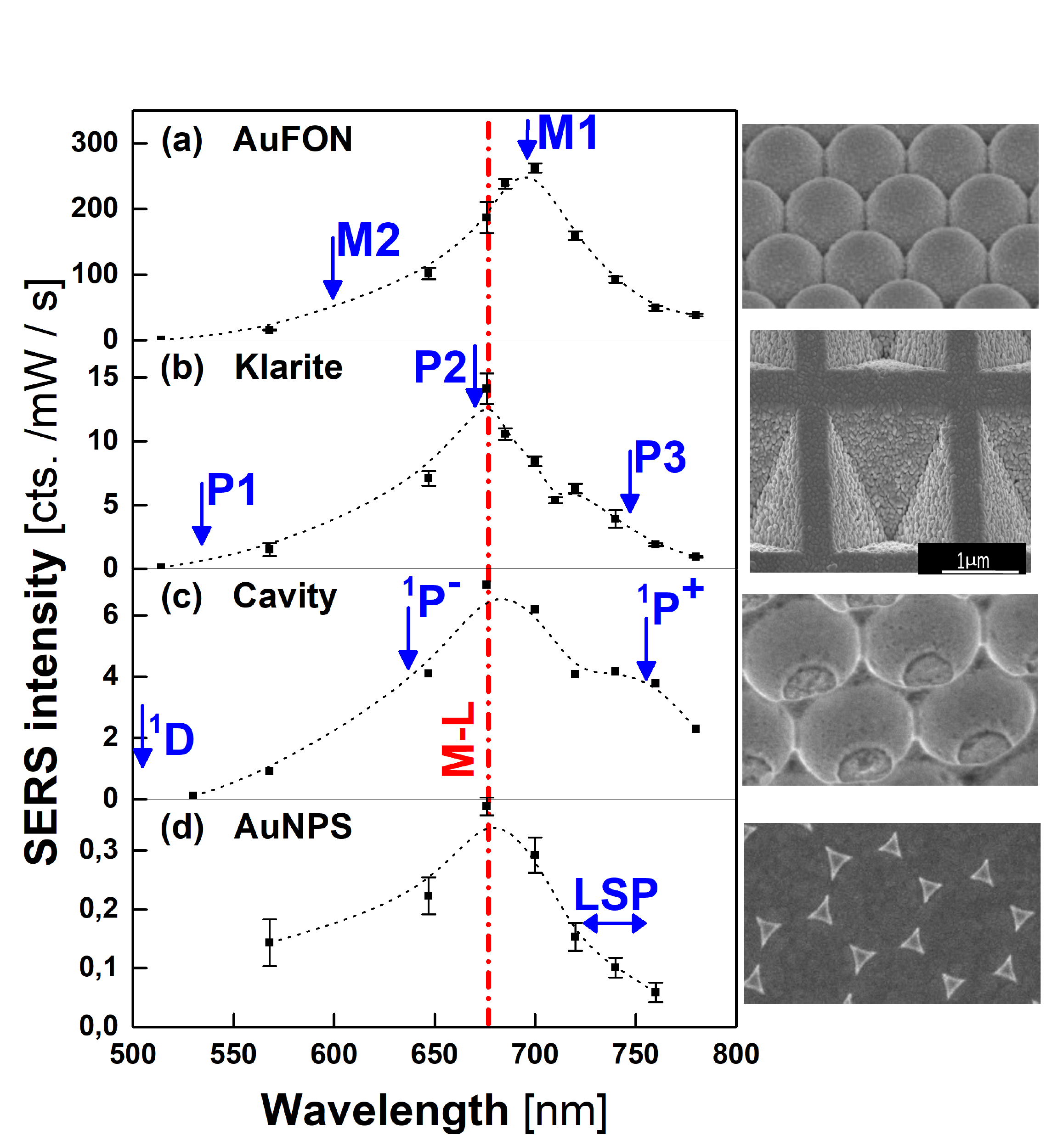}
\caption{Panels a-d present a comparative study of SERS resonant scans obtained under the same conditions and using the same molecular probe as the experiments in Figs.\ref{Fig02} and \ref{Fig03}, for AuFON, square pyramidal Au covered pits (commercial substrate Klarite), Au segment sphere void cavities, and Au triangular nanoparticles, respectively. SEM images of the studied substrates are shown at the right of each panel. AuFON (a), cavities (c) and Au NPs (d) were all fabricated using nanosphere lithography methods with 500~nm polystyrene spheres. The labels in each panel identify the plasmon modes determined from reflectivity measurements equivalent to those presented for the MFON substrates in Figs.\ref{Fig02} and \ref{Fig03}.} 
\label{Fig04}
\end{figure}

Figure~\ref{Fig04} presents a comparative resonant SERS study of a diverse variety of ordered Au plasmonic substrates, namely AuFON (a), square pyramidal pits of the commercial Klarite substrate (b)~\cite{Perney2007}, segment sphere void cavities (c)~\cite{Tognalli2012,Kelf2006,Mahajan2007,Mahajan2009}, and triangular nanoparticles (d).~\cite{Jensen1999} Each panel in Fig.\ref{Fig04} presents, at its right, a SEM image of the structure. Substrates (a), (c) and (d) were all fabricated by nanosphere lithography using 500~nm polystyrene spheres. As rugosity is known to have a potentially relevant effect on the SERS efficiency~\cite{GuerraHernandez2015,Fernandez2009,Trugler2014}, it was determined for all substrates (excluding the triangular Au NPs that are not extended and thus are not directly comparable) using AFM scans finding comparatively similar values in all cases. Grain sizes span from 0 to ~ 120~nm with a peaked distribution with maxima around 40-60~nm, with similar degree of roughness for the cavities and MFON substrates and slightly lower frequency of occurrence of the grains for the commercial Klarite substrate.  The extended substrates were also monitored for their homogeneity in terms of Raman efficiency, finding on $5 \times 5 \mu$m$^2$ square areas dispersions of around 17\%, 21\%, and 31\% for the AuFON, Klarite, and cavity structures, respectively. Assuming then that constructively these substrates are comparable, we extract two important conclusions. First, again and in all four cases, the M-L transition seems to be determinant to the spectral position where maximum SERS efficiency is found. And second, by quite far (a factor of at least 20) the AuFON substrate results in a much larger (and more homogeneous) Raman efficiency. For comparison, we note here that experiments performed on identical 4-MBA self-assembled monolayers on flat (non-structured) Au substrates lead to no observable Raman signals even with 20 times larger acquisition times. To understand the physics behind the observed larger electromagnetic SERS enhancement for the MFON substrates, and to identify the origin of the M1-M3 plasmon modes and their contrasting efficiency for SERS, we briefly describe next its modeling based on finite element methods.

%=================================
%=================================

\begin{figure}[hbt!]
\centering
\includegraphics[width=0.7\textwidth]{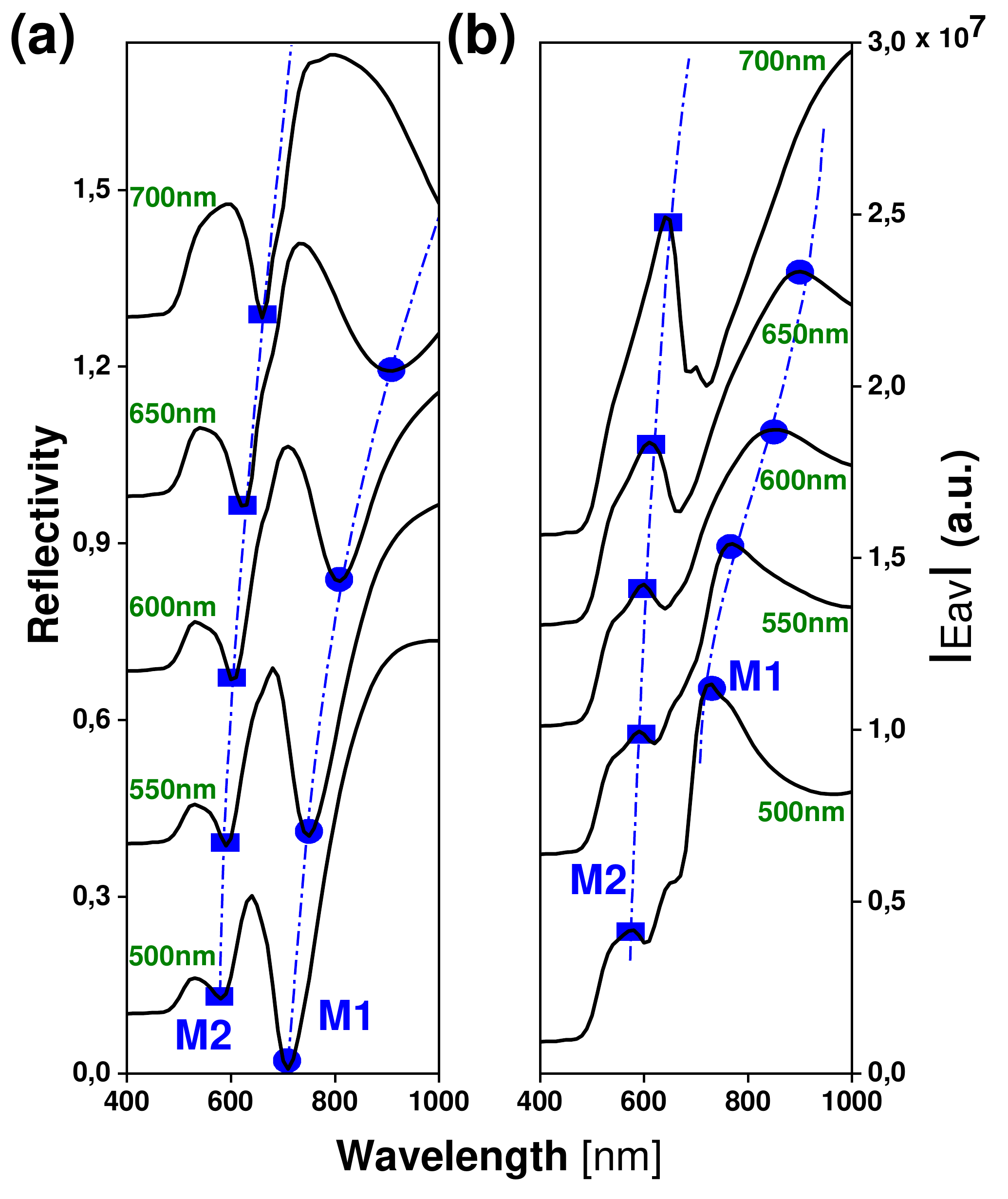}
\caption{Calculated wavelength dependence of the far-field reflectivity (a) and near-field amplitude (b) for a plane wave incident on AuFON substrates of varying sphere size. These substrates are modeled as solid Au half spheres on a planar  and uniform Au film. The solid connected symbols identify the plasmon modes.} 
\label{Fig05}
\end{figure}

\subsection{Theoretical modeling of metal-film over nanosphere substrates}

We evaluate the electromagnetic response of the nanostructured surface of a AuFON substrate when it is excited by a plane wave by solving for the full-field Maxwell equations in 3D using the finite element method. The standard dielectric function of Au is used as given in Ref.~\citenum{Johnson1972}. Periodic boundary conditions are incorporated by ensuring perfect agreement in the triangular mesh within each of the three pairs of partner faces defining the hexagonal arrangement. The computed scattering matrix includes all allowed diffraction modes obtained using the customary Ewald criteria for the working plane-wave wavelength and incident direction. To first test the qualitative behavior of the mode, we perform calculations for an hexagonal arrangement of solid Au semi-spheres over a uniform Au surface. This is shown in Fig.\ref{Fig05}, were both the far field reflectivity (a) and the near field averaged on the surface (b) are shown as a function of wavelength and for increasing sphere size (from bottom to top). The magnitude of the near field determines the electromagnetic enhancement affecting the SERS efficiency, and thus the two shown panels can be related to the similar ones in Figs.\ref{Fig02}. The agreement is good in several features, namely: i) two main plasmon modes are observed in the relevant wavelength range, ii) these disperse to larger wavelengths for increasing nanosphere size, and iii) the smaller energy mode, identified as M1, leads to the largest averaged near field and thus is expected to provide the stronger Raman signals as experimentally observed.

\begin{figure}[hbt!]
\centering
\includegraphics[width=0.7\textwidth]{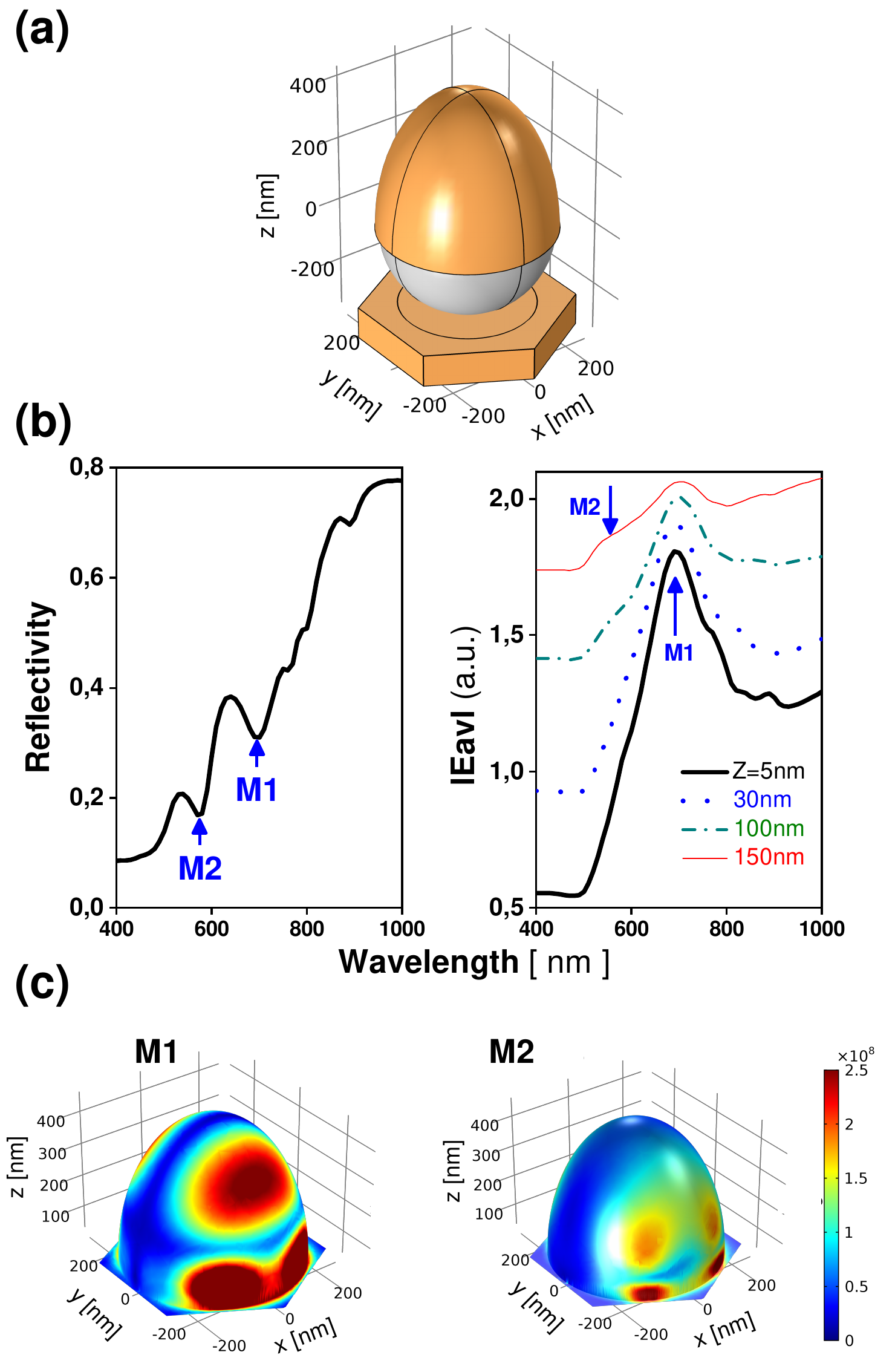}
\caption{(a)Scheme of the ellipsoid shape of the deposited Au film, internal polystrene sphere, and metal Au base used in the calculations.(b) Wavelength dependence of the calculated far field reflectivity (left panel) and  associated averaged magnitude of the surface  near-field, for an AuFON substrate of 500~nm spheres and 180~nm film thickness. In the right panel different curves are presented for varying distance from the surface. M1 and M2 identify the plasmon resonances. (c) Color maps of the spatial distribution of the magnitude of the electromagnetic fields for plasmon modes M1 and M2, with incident polarization along $x$.}
\label{Fig06}
\end{figure}

While the qualitative agreement between theory and experiment is reasonably good with such a simplified description of the substrates, we have observed that the quantitative determination of the plasmon energies and size dependence is strongly sensitive on the details of the structure. Calculations with a more realistic description of the AuFON substrates are presented in Fig.\ref{Fig06}, where the polystyrene spheres have been included (as a dielectric material of index of refraction $n=1.59$), and the film cover was allowed to have an ellipsoid shape to account for a larger deposit of metal on the top (see a scheme in Fig.\ref{Fig06}(a)). For this simulations we use 500~nm diameter spheres, the Au thickness was taken as 180~nm at the top of the sphere, and light was incident at angles $\theta$=25$^\circ$ and $\phi$=0$^\circ$. The interstitial spaces between spheres have been simulated with and without pyramidal Au arrangements, without significant variations. In this case excellent agreement with the experimental results is obtained as shown in Fig~\ref{Fig06}(b), both in the spectral position of the modes and the relative magnitude of the averaged surface near-field associated to the M1 and M2 plasmon modes. The strong sensitivity of the latter on the distance to the surface of the Au film is illustrated on the right panel of Fig.\ref{Fig06}(b). The color maps in Fig.\ref{Fig06}(c) further clarify the contrasting response of plasmons M1 and M2. The M1 one mode has very strong associated electromagnetic fields very close to the surface and precisely in between neighbor spheres as known to be relevant for nanoparticle dimers. This result also explains the comparative high efficiency of the MFOM substrates when contrasted with the square pyramidal pit and segment sphere void cavities as shown in Fig.\ref{Fig04}: the latter concentrate important parts of the fields in the cavity's empty space~\cite{Tognalli2012,Kelf2006,Mahajan2007,Mahajan2009}, while molecules are immobilized on the surface.

%%%%%%=================================
%%%%%%=================================
\section{Conclusions}

We have reported a comprehensive comparative study of ordered SERS plasmonic substrates fabricated using nanosphere lithography and, to the best of our knowledge, the first full resonant study in which both the laser wavelength and the plasmon resonances are independently tuned to provide a complete experimental description of the resonant processes at play. This was done relying on one of the standard (intrinsically non-resonant) probes used for Raman enhancement studies, namely a self-assembled covalently bound monolayer of 4-mercaptobenzoic acid (4-MBA) that is known to form an ordered and densely packed monolayer on the surface of the metal. The concluding answer to the posed question, ``has the chemical contribution a secondary role in SERS?'' is clearly no.

We conclude that the first and undoubtedly most important resonance is the surface plasmon resonance of the metallic substrate (the electromagnetic mechanism). In fact, the absence of nanostructuring of the metallic surfaces leads to no observable Raman signals. However, another resonance of the system has also a key role in the enhancement, namely the charge-transfer between the molecule and the Fermi level of the metal (usually termed as the chemical mechanism). It has become experimentally clear that in order to adequately explain the SERS enhancement, the combined molecule-metal system has to be considered.\cite{Lombardi2017}

Our experiments demonstrate that resonant experiments are indispensable to fully address the SERS mechanisms involved, but most importantly that even these might not be enough.  It becomes clear that experiments done with a single laser and tuning the plasmon energies for resonance might be misleading. And the reversed alternative, i.e. tuning the laser for a fixed plasmonic substrate, can also provide an incomplete and potentially ambiguous picture. It is to be expected that most experimental parameter that can be varied to probe a system will have an influence via both mechanisms, electromagnetic and chemical, making the separation of effects difficult.\cite{Campion1998} 

Overall, the emerging picture calls for a unified description of Raman resonances in SERS, as e.g., previously introduced in Ref.~\citenum{Londero2013}. Coincident with these discussed unified models, our experiments evidence that in the studied substrates the plasmon transitions donate intensity to charge-transfer transitions, and also suggest that the plasmonic resonances are sufficiently broad to provide enhancement over a large wavelength range, implying that they are not primarily responsible for the observed spectral features. In agreement with these unified theories, our experiments indicate that both electromagnetic and so-called chemical resonances are coupled and thus the resonant denominators cannot be divided out  to consider the contributions separately. 

From our comparative study of ordered plasmonic substrates our conclusion is that, to fully exploit the electromagnetic enhancement for the detection of surface-deposited molecules, convex substrates (as MFON) seem to be much more efficient than concave ones (as the studied sphere segment void cavities and the square pyramidal pits). And probably much more important than that, we conclude that it is extremely important to take into account and understand the so-called chemical mechanism for both fundamental reasons and for its relevance to analytical applications. As previously alerted, since the resonant effects are multiplicative, unexpected chemical enhancement could lead to analytical conclusions which are quantitatively wrong.\cite{Campion1998}

\paragraph*{\textbf{Funding.}} 
The authors acknowledge financial support from the ANPCyT (Argentina) under grants PICT-2018-03255 and PICT-2020-SERIEA-03123. A.A.R. aknowledges support by PAIDI 2020 Project No. P20-00548 with FEDER funds.

\paragraph*{\textbf{Disclosures.}} The authors declare that there are no conflicts of interests related to this article.

\paragraph*{\textbf{Data availability.}} 
Data underlying the results presented in this paper are available from the corresponding author upon reasonable
request.

%\begin{suppinfo}

%\end{suppinfo}

\mciteErrorOnUnknownfalse

%\bibliography{references}
%\newpage

%\section*{For Table of Contents Use Only}
%\emph{Title}: Wavelength dependence of SERS in Au-film over nanosphere substrates\\[0.8em]
%\emph{Authors}: Luis A. Guerra Hern\'andez, Andr\'es Reynoso, {Alejandro Fainstein}.\\[0.8em]
%\emph{Description}: describe TOC.............-----------
%.\\[0.8em]

%\centering\includegraphics[height=2.375in]{TOC}
%\centering\includegraphics[height=8cm]{TOC}
\end{document}